\documentclass[apj]{emulateapj}

\usepackage{times}

\shorttitle{Brooks, Reep, \& Warren}
\shortauthors{Brooks, Reep, \& Warren}

\begin{document}


\title{Properties and Modeling of Unresolved Fine Structure Loops Observed in the solar 
transition region by \textit{IRIS}}

\author{David H. Brooks\altaffilmark{1,3}, Jeffrey W. Reep\altaffilmark{2,4},
  and Harry P. Warren\altaffilmark{2}}

\affiliation{\altaffilmark{1}College of Science, George Mason University, 4400 University Drive,
  Fairfax, VA 22030, USA}

\affiliation{\altaffilmark{2}Space Science Division, Naval Research Laboratory, Washington, DC
  20375, USA}

\altaffiltext{3}{Current address: Hinode Team, ISAS/JAXA, 3-1-1 Yoshinodai, Chuo-ku, Sagamihara,
  Kanagawa 252-5210, Japan.}

\altaffiltext{4}{National Research Council Postdoctoral Fellow.}


\begin{abstract}
Recent observations from the Interface Region Imaging Spectrograph (\textit{IRIS}) have discovered
a new class of numerous low-lying dynamic loop structures, and it has been argued that they are the
long-postulated unresolved fine structures (UFS) that dominate the emission of the solar transition
region. In this letter, we combine \textit{IRIS} measurements of the properties of a sample of 108
UFS (intensities, lengths, widths, lifetimes) with 1-D non-equilibrium ionization simulations using
the HYDRAD hydrodynamic model to examine whether the UFS are now truly spatially resolved in the
sense of being individual structures rather than composed of multiple magnetic threads. We find
that a simulation of an impulsively heated single strand can reproduce most of the observed
properties suggesting that the UFS may be resolved, and the distribution of UFS widths implies that
they are structured on a spatial scale of 133\,km\, on average.  Spatial scales of a few hundred km
appear to be typical for a range of chromospheric and coronal structures, and we conjecture that
this could be an important clue to the coronal heating process.
\end{abstract}



\section{introduction}

From Skylab data analysis, \citet{feldman_1983} identified several discrepancies between
observations of the emission from the transition region, which he defined as the solar plasma in
the 0.02--1.0\,MK temperature range, and the predictions of classical models of the interface
connecting the chromosphere and corona. First, the emission above the solar limb comes from more
extended heights than the thin transition layer expected from one continuous structure. Second,
there is a large discrepancy between the observed emission at 0.25\,MK and the predictions of
theoretical models \citep{athay_1982}. \citet{feldman_1983} argued that these observations,
together with other evidence based on off-limb line width measurements and electron
densities in different solar regions, suggest that most of the transition region emission
originates in unresolved fine structures (UFS) that are magnetically isolated from the chromosphere
and corona. Subsequently, \citet{feldman_1987}, \citet{feldman_1998}, and \citet{feldman_etal2001}
presented further evidence for this view. In particular, \citet{feldman_1998} showed 
that the plasma composition enhancement at transition region temperatures is different than
that at coronal temperatures, supporting the idea that the UFS are disconnected from the corona,
and have relatively shorter lifetimes than coronal structures.

This suggestion led to theoretical developments such as the ``cool loop'' model of \citet{antiochos&noci_1986}. The growing awareness of the
temporal variability of the solar atmosphere, however, led others to question the existence of UFS based on classical models of the transition region modified
to incorporate dynamic effects \citep{wikstol_etal1998}. Without instruments with sufficient spatial resolution to settle the debate, however, these diverging
views were not reconciled.

The launch of the Interface Region Imaging Spectrograph \citep[\textit{IRIS},][]{depontieu_etal2014}, however, has provided a new opportunity to 
observe the transition region with high spatial resolution and high cadence. Using \textit{IRIS}, \citet{hansteen_etal2014} discovered numerous rapidly-varying low-lying loops
at transition region temperatures and argued that these are the UFS, whose existence was predicted by \citet{feldman_1983}. 
\citet{hansteen_etal2014}
found that these loops are highly dynamic; often showing red and/or blue shifted velocities of 80\,km s$^{-1}$ or more. 
They have half lengths of 2--6\,Mm, reach heights of 1--4.5\,Mm and are about three times brighter on average than spicules. They light up
in segments and are short-lived, though systems of several may persist for tens of minutes. Since they have only 
recently been detected little else is known about their properties, and it
has not been demonstrated that they are the dominant contributors to the transition region emission. That could
instead come from spicules \citep{depontieu_etal2011}, and their relative contributions remain to be quantified.
Throughout this work, however, we have followed the argument of \citet{hansteen_etal2014} and refer to the UV loops studied here
as UFS. 

\begin{figure*}
  \centerline{\includegraphics[width=0.95\linewidth,viewport=80 345 535 565]{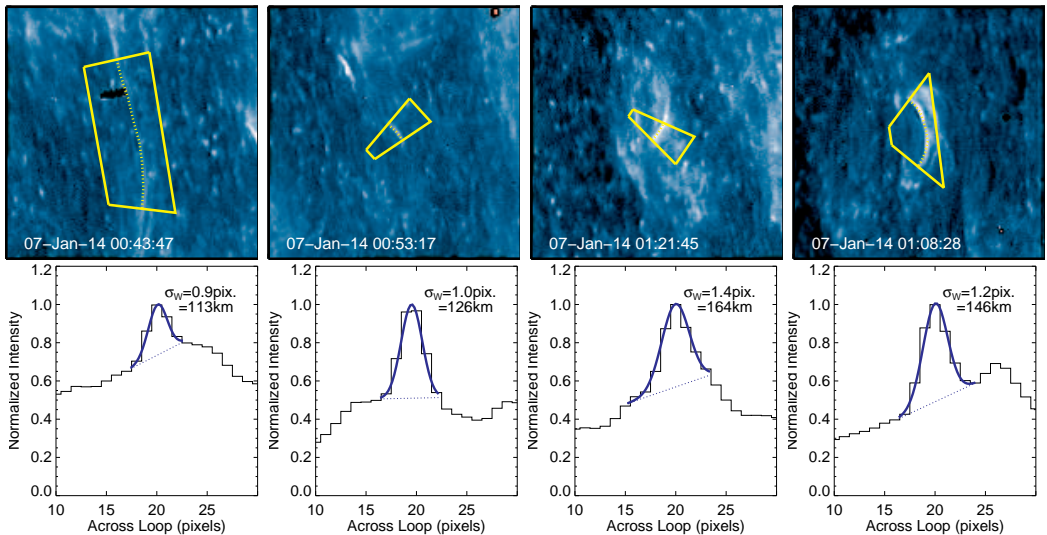}}
  \vspace{-0.1in}
\caption{Example UFS segments. Top row: \textit{IRIS} slit-jaw images (SJI) with the loop segments
  highlighted in yellow. These images have been sharpened with a Gaussian filter. Bottom row: cross-field normalized intensity profiles (solid histogram) with
  Gaussian fits (blue solid line) and backgrounds (blue dotted line) overlaid. The Gaussian width
  in \textit{IRIS} pixels and km is shown in the legend. The interpolated data have been resampled
  to show the instrument pixel scale.
\label{fig:fig1}}
\end{figure*}

\textit{IRIS} has the sensitivity and
resolution to detect the UFS, 
but the question arises as to whether that sensitivity and resolution is sufficient to {\it spatially
 resolve} them as single monolithic structures, or whether they are bundles of
sub-resolution magnetic strands that remain {\it spatially unresolved}. This point was raised by
\citet{hansteen_etal2014}, who noted that UFS that appear as monolithic structures in \textit{IRIS}
observations show substructure in their numerical simulations, and is a question of recent interest
because the answer defines the requirements for future solar instrumentation that we hope can
determine the true properties of solar atmospheric structures. It also guides our theoretical
thinking, because many chromospheric and coronal heating mechanisms are expected to release energy
on very small spatial scales ($\sim 10$'s of meters). Physical models of observations of coronal
loops from \textit{Hinode}, \textit{SDO}, and \textit{Hi-C}, however, suggest that we are already
close to resolving them with current instruments, and that they have spatial scales of a few
hundred km \citep{brooks_etal2012,brooks_etal2013}. A few hundred km seems to be a common scale for
many solar atmospheric structures from Type II spicules \citep{pereira_etal2012} to ``coronal
rain'' condensations \citep{antolin&vandervoort_2012}, flaring and post-flare loops
\citep{cirtain_etal2013,jing_etal2016}, and even prominence threads \citep{okamoto_etal2007}. So,
if correct, the challenge becomes how to convert the small scale energy release into this typical
preferred size.

Here we present new measurements of the spatial scales of the UFS. We also use the
observed lifetimes, half-lengths, and peak intensities as input to numerical hydrodynamic
simulations of the cooling times of the UFS. By comparing the model with the \textit{IRIS} data, we
conclude that the observed lifetimes are consistent with a simulation of a single thread,
suggesting that they may be spatially resolved.

\section{observations}

\begin{figure*}
  \centerline{\includegraphics[width=0.95\linewidth,viewport=100 380 520 540]{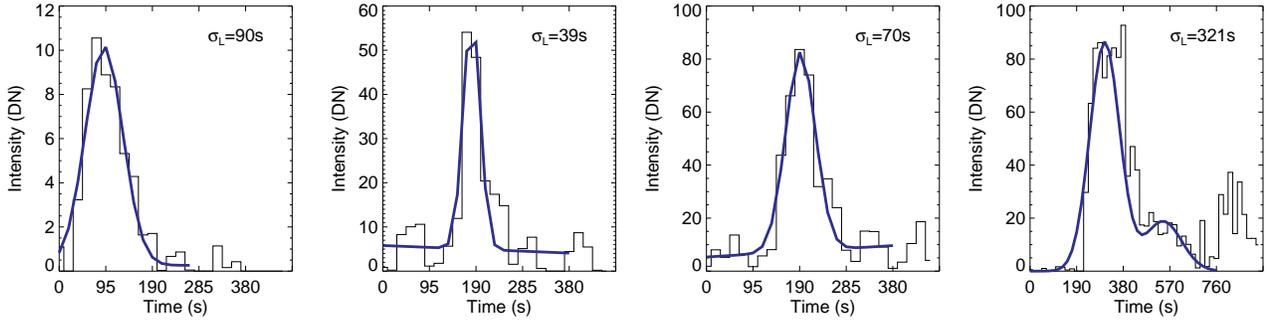}}
  \vspace{-0.25in}
\caption{Light curves for the UFS in Figure \ref{fig:fig1}.
We show the extracted intensities (histogram) with Gaussian fits (blue line)
overlaid. The FWHM of the Gaussian fit for each case is given in the legend.
\label{fig:fig2}}
\end{figure*}

\begin{figure*}[t!]
\centerline{\includegraphics[viewport= 110 500 510 790,clip,width=0.75\textwidth]{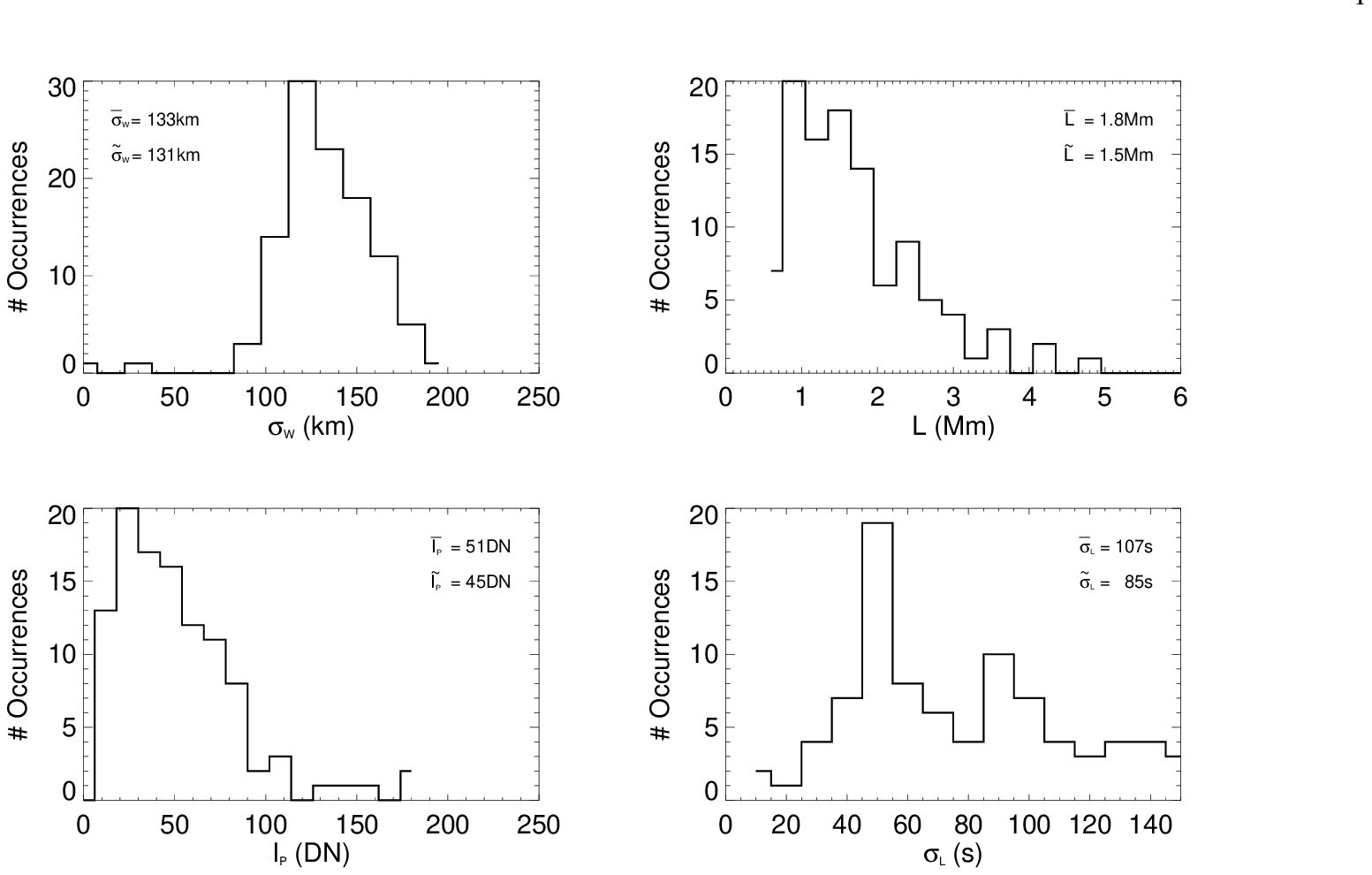}}
\caption{Histograms displaying our analysis results for the sample of UFS. 
Top left: Distribution of cross-field (Gaussian) widths, $\sigma_W$, in km. Top right: Distribution of 
half-lengths, $L$, in Mm. Bottom left: Distribution of peak intensities, I$_p$, in DN. Bottom right: 
Distribution of UFS FWHM lifetimes, $\sigma_L$, in s. The mean and median values are indicated in the legend. 
\label{fig:fig3}}
\end{figure*}

\citet{depontieu_etal2014} describe \textit{IRIS} in detail.
Here we only use \ion{Si}{4} 1400\,\AA\, slit-jaw images. 
The data were obtained between 2013 December 9 and 2014 January 7 using the \textit{IRIS} observing ID
3800259453, which takes 8s exposures of a 120$''$ by 129$''$ field-of-view and was run near the solar limb 
where it is relatively easier to identify UFS. The spatial pixel size is 0.167\arcsec. The data have been processed, calibrated, and coaligned to level-2 and were obtained
from the search facility at Lockheed Martin. 

We examined $\sim$\,5 hours of observations and visually identified 108 UFS. 
Many of them brighten only partially, or the brightenings move along the loop, or
the UFS themselves move rapidly and change dimensions. 
So there may be some selection bias towards UFS that are visually prominent and relatively
easier to isolate from surrounding features. During our study, however, our analysis techniques were developed 
to better handle structures that are 
moving or changing shape.

To measure the UFS intensities and widths we used the same procedure as in our earlier work on coronal loop properties 
\citep{warren_etal2008,brooks_etal2012,brooks_etal2013}. The method is based on the work
of \citet{aschwanden_etal2008}, and extracts the cross-UFS intensity profile by interpolating along the structure within
a selected segment, straightening it, and averaging the intensities along the UFS axis. A first-order polynomial is then
fit to the background between two selected positions in the intensity profile, and a Gaussian fit is made to the background
subtracted profile. The UFS intensity and width ($\sigma_W$) are then the area and width of the Gaussian, respectively. 
Examples are shown in Figure \ref{fig:fig1}.

The same segment, UFS axis, and background definitions are then used to extract the intensities for every image in the data cube 
in order to prepare a light curve for measuring the lifetimes. As mentioned, there is a problem here if the UFS
are moving or changing dimensions, because the selected axis may not be co-located with the structure at all times. This issue
also sometimes occurs for UFS that are long-lived because the \textit{IRIS} coalignment can be affected by 
spacecraft pointing drift while observing the limb. The automatic coalignment procedure and post-processing by cross-correlation
do not always fully correct
for this because they tend to lock on to the slit portion and fiducial marks in the slit-jaw images.
To mitigate this problem, we allow for dynamic movement
of the selected background positions in the cross-UFS intensity profile. Since these movements are always smaller than the 
range of the perpendicular segment, the averaged profile includes the UFS even if it has moved, and the dynamic movement
of the background positions ensures that the extracted intensity and width are correct and reflect what is observed in the images. 

Example UFS light curves are shown in Figure \ref{fig:fig2}. 
We fit a polynomial background and Gaussian function to the light curve of each UFS, and take the 
full width at half maximum (FWHM) as the lifetime ($\sigma_L$). The peak intensity ($I_P$) of the Gaussian is also recorded
for each UFS. Note that the degree of the background polynomial used in the fit is dependent on how prominent the UFS light 
curve is above the background. Also, sometimes the UFS brightens and fades but remains visible for some time. In these cases 
we fit multiple Gaussians to the light curve and sum the FWHMs to get the lifetime. The fourth panel in Figure \ref{fig:fig2}
shows an example of such a case.
Finally, we also measure the UFS half-length ($L$) by extracting the cross-UFS intensity profile averaged along the half-length
of the structure, straightening it, and measuring the straightened length. 

The results of our analysis are shown as histograms in Figure \ref{fig:fig3}. 
The UFS widths fall in the range 35--192\,km with an average of 133\,km; 
the UFS half-lengths fall in the range 0.6--9.5\,Mm with an average of 1.8\,Mm;
the UFS peak intensities fall in the range 9--182\,DN with an average of 51\,DN;
and the UFS lifetimes fall in the range 5--323\,s with an average of 107\,s.
These peak intensities and lifetimes are consistent with the results of \citet{hansteen_etal2014}. Our half-lengths are towards,
and smaller than, the lower range of their values, but this can likely be explained by the different analysis methods. 
\citet{hansteen_etal2014} do not quote any width measurements.

We note that the exact nature of these distributions is not clear from our limited sample size. It could be
that they are actually power-law distributions, in which case the means would not be defined. 

\section{simulations}

\begin{figure*}
\centerline{\includegraphics[viewport= 20 370 580 540,clip,width=0.95\textwidth]{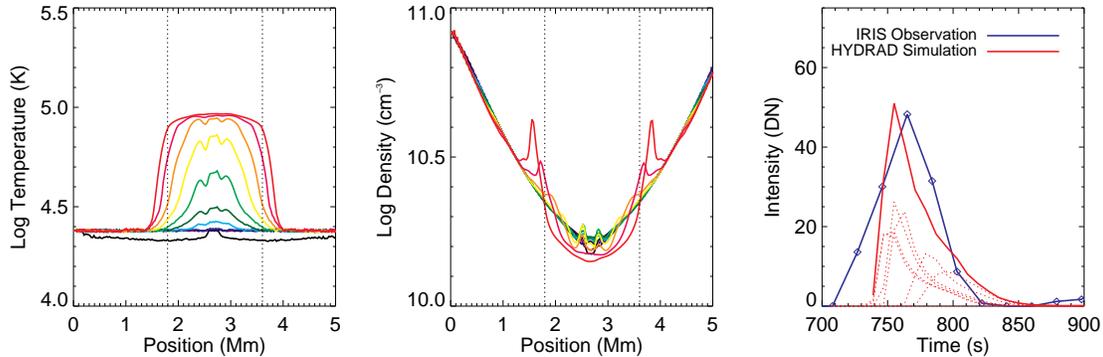}}
\caption{Simulation of the properties of one of the UFS in our sample. This UFS has a half-length
of 0.9\,Mm, a peak intensity of 47.1\,DN and a lifetime of 50.7\,s. Left panel: The
distribution of temperature along the loop during the simulation.
Middle panel: the distribution of density. The temperature and
density distributions are plotted every 5\,s going from blue to red. The vertical dotted lines show
that a loop of the required half-length (0.9\,Mm) is formed. Right panel: Comparison
between the simulated (red) and observed (blue) intensities for the same UFS. The total intensity of the half-loop
is shown with the solid red line. The intensities of a subset of the individual pixels are shown with dotted red lines. Only
half of the individual pixels are shown for clarity.
\label{fig:fig4}}
\end{figure*}

\begin{figure*}
\centerline{\includegraphics[viewport= 20 350 580 540,clip,width=0.95\textwidth]{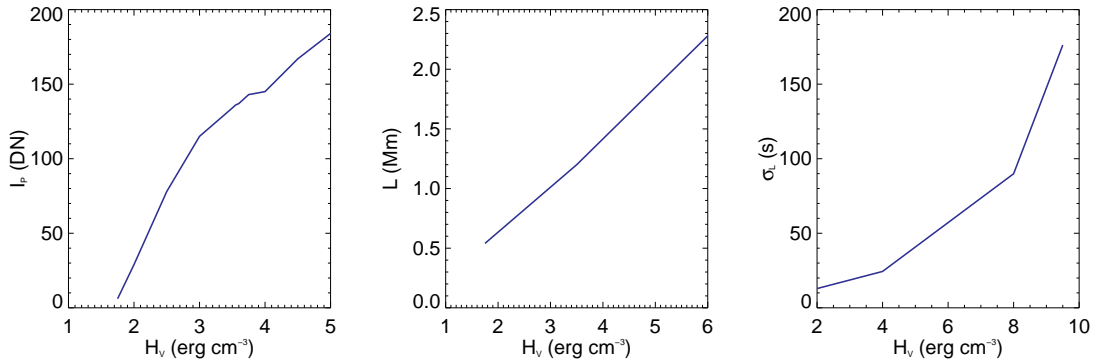}}
\caption{Summary plots of the simulated UFS peak intensity, I$_p$, half-length, $L$, and lifetime, $\sigma_L$, as a function of the volumetric heating, H$_V$.
\label{fig:fig5}}
\end{figure*}

We have run simulations with the HYDrodynamics and RADiation code (HYDRAD; \citealt{bradshaw&mason_2003,bradshaw&cargill_2013}), which solves the hydrodynamic equations appropriate to a two-fluid plasma confined to an isolated, one-dimensional magnetic flux tube.  We treat radiative losses with a full calculation, using emissivities from CHIANTI v.8 \citep{dere_etal1997,delzanna_etal2015}, assuming photospheric abundances \citep{asplund_etal2005}, and allow for non-equilibrium ionization of hydrogen and silicon.  We calculate the \ion{Si}{4} emission line following the methodology of \citet{bradshaw&klimchuk_2011}, using the \textit{IRIS} instrumental response from \verb+iris_get_response+ in SSW.  

We assume that the loop begins with a flat temperature profile of 24 kK, {\it i.e.} that the loop has not been heated to coronal temperatures initially and that there is no background heating term.  We then heat the loop uniformly, with a triangular temporal profile, for a given duration and maximum heating rate.  As the loop evolves, we calculate the \ion{Si}{4} emission and test for consistency with the observed values of intensity, lifetime, and loop length.

Figure \ref{fig:fig4} shows a simulation that reproduces the properties of one of the UFS in the sample. This UFS has a half-length
of 0.9\,Mm, a peak intensity of 47.1\,DN and a lifetime of 50.7\,s. The volumetric heating
rate in the simulation is 6$\times$10$^{-2}$ erg cm$^{-3}$ s$^{-1}$ and the heating duration is 100\,s. The temperature increases
to $\log$T = 4.96 and the density decreases to $\log$N$_e$ = 10.15. This forms a loop of the required 
average half-length (0.9\,Mm) emitting very close to the temperature of the peak of the 
\ion{Si}{4} 1393.755\,\AA\, contribution function ($\log$T = 4.9), which is the strongest \ion{Si}{4} line in the SJI passband.

Figure \ref{fig:fig4} also compares the totaled intensity over the loop as a function of time for the simulation with 
the light-curve of the same UFS. 
The simulated \ion{Si}{4} 1393.755\,\AA\, emission
has been scaled down by a factor of 16, in this example only, to match the peak intensity because the line-of-sight depth is an unknown parameter
in the simulation. More importantly, the 
observed intensity profile is modeled quite well and a Gaussian fit to the simulated intensity profile gives a 
lifetime of 43\,s, which is within 20\% of the observed value.

We also plot the contributions to the total intensity from several individual pixels in the simulation (dotted lines). 
Interestingly, these are offset in time and have different peak brightnesses. 
This reproduces another aspect of the observations that was noted by \citet{hansteen_etal2014} and
discussed earlier: that the UFS light up in segments with different intensities at different times.  

We found that the model can also reproduce most of the properties of the entire sample of UFS if we modify the volumetric
heating rate or heating duration. Figure \ref{fig:fig5} illustrates this by showing the peak intensity, half-length, and 
lifetime of the simulated UFS as a function of, H$_V$, the volumetric heating (product of rate and duration). The range
of values of H$_V$ shown produce peak intensities of 6.1--184\,DN, half-lengths of 0.5--2.3\,Mm, and lifetimes of 13--176\,s. 
These simulated ranges cover 99\% of the peak intensities in our UFS sample, 76\% of the half-lengths,
and most significantly, 81\% of the lifetimes. 
It is likely that adjusting the volumetric heating higher or lower by finding a new combination of rate and duration
should reproduce the remaining observed properties. 

\citet{hansteen_etal2014} commented that their magnetohydrodynamic modeling predicts that short low-lying loops seldom reach temperatures
higher than $\log$T = 5, and would be difficult to observe, e.g, with the Atmospheric Imaging Assembly \citep[\textit{AIA},][]{lemen_etal2012} on the Solar Dynamics Observatory \citep[\textit{SDO},][]{pesnell_etal2012}, 
due to insufficient spatial resolution and absorption of EUV emission in the 171\,\AA\, and 193\,\AA\, pass-bands.
For a large range of energies the loops in our hydrodynamic model also do not heat higher than this temperature,
and the simulation predicts no emission in 
higher temperature coronal images such as 171\,\AA\, or 193\,\AA. This is because the radiative loss function peaks close
to this temperature ($\log$T$\sim$5.4), so a large amount of energy is required to overcome the losses.
We have also independently looked at the AIA data for 
several cases and found no clear emission. Conversely, these pass-bands also contain transition region
lines and our model does predict emission from them of $\sim$30\,DN pixel$^{-1}$ s$^{-1}$. Examining several
examples, we found that the background emission in the 171\,\AA\, and 193\,\AA\, channels around the UFS location
is at least 135\,DN and 124\,DN, respectively, so any transition region emission is swamped by this background.

Note that we have not attempted to reproduce all the UFS properties simultaneously for all 108 UFS in the sample, 
which would be a very large undertaking. By exploring parameters in the model, however, we found that for the longest
duration heating, the plasma exceeds $\log$T = 5 and the peak intensity in the \ion{Si}{4} emission no longer
corresponds to the maximum loop length since that occurs at higher temperatures. These cases appear more like traditional
coronal loops, but produce only very short loops ($<$0.4\,Mm) when they are emitting in \ion{Si}{4}. In fact,
it is difficult to produce very long \ion{Si}{4} loops with extended heating: a minority ($\sim$27\%) of the UFS
whose lifetimes are reproduced by our simulations are longer than predicted, indicating that there is a 
discrepancy between the modeling and observations for the longer duration UFS that needs further investigation. For the
objectives of this letter, however, we stress that the loop length is not a property that can be modified by adding
more strands. 

\section{Summary and discussion}

We have examined the properties of a sample of 108 UFS using \textit{IRIS} \ion{Si}{4} SJI images, 
and used them to guide numerical hydrodynamic simulations.
We find that in the majority of cases,
the lifetimes, half-lengths, and peak intensities can be reproduced by a single
thread model, suggesting that they may be spatially resolved.

Another interesting aspect of the modeling is that it seems to naturally explain why the UFS
light up in segments with different brightnesses at different times, but our
most significant result is that the observed UFS lifetimes can be matched with a single 
thread model. This has proven difficult to
achieve with `warm' coronal loops, because they are observed primarily in the cooling phase 
\citep{ugarteurra_etal2009}, and a model that heats them to an equilibrium at 1\,MK\, does
not produce sufficiently high densities \citep{winebarger_etal2003}. This has been one of the primary drivers of the development
of the multi-thread model of coronal loops. The UFS, in contrast, are observed in the heating
phase, and so their lifetimes can be reproduced by bringing a single strand to a temperature of $\sim$0.5\,MK\,
and sustaining it. Our modeling shows that densities in excess of $\log$N = 10 can be achieved.

We also measured the UFS widths and, if they are truly resolved, the distribution shows 
that they are structured on a spatial scale of 133\,km\, on average. As noted earlier,
a few hundred km seems to be a common spatial scale for many structures in the solar atmosphere and is much
larger than current theoretical modeling predicts. Even if the heating mechanism itself operates
on much smaller spatial scales, the plasma seems to respond to heating with coherence and collective
behavior on these characteristic scales, and understanding why is becoming an important question.

We speculate that this observation tends to favor a coronal heating mechanism with some threshold as an
onset condition. For example, when observing the cross-field intensity profile of a coronal loop, the 
atomic physics is assumed to be known, and for a single thread model the emission measure relates the loop
radius to the electron density \citep{brooks_etal2012}. The density depends on the magnitude of the heating
\citep{klimchuk_2006}, which in turn depends in an unknown way on the released magnetic energy. In some theories of
coronal heating, the amount of magnetic energy available can be related to a property that switches on
above a minimum threshold. In the nanoflare reconnection concept, for example, the Poynting flux into 
the corona that stresses the magnetic field can be related to the mis-alignment angle between the
horizontal and vertical components of the field \citep{klimchuk_2006}. \citet{dahlburg_etal2005} have
argued that the secondary instability occurs when this mis-alignment (shear) between the fields (current
sheets) exceeds a threshold angle. In this example, our idea is that this minimum threshold angle leads to 
a characteristic Poynting flux, heating, density, and therefore loop radius. If the instability did not 
have this switch-on property, then it could presumably occur for any angle, and produce any range of 
energies, densities, and spatial scales.


\acknowledgments 

DHB thanks Viggo Hansteen for \textit{IRIS} data analysis advice.
The work of DHB and HPW was performed under contract to the Naval Research Laboratory and was
funded by the NASA \textit{Hinode} program.  This research was performed while JWR held an NRC
Research Associateship award at the US Naval Research Laboratory with the support of NASA.  IRIS is
a NASA small explorer mission developed and operated by LMSAL with mission operations executed at
NASA Ames Research center and major contributions to downlink communications funded by ESA and the
Norwegian Space Centre.  CHIANTI is a collaborative project involving George Mason University, the
University of Michigan (USA) and the University of Cambridge (UK).


\bibliography{ms}
\bibliographystyle{ms}

\end{document}